\documentclass[twocolumn,prb]{revtex4}
\usepackage{amsfonts}
\usepackage[T1]{fontenc}
\usepackage{amsmath,amsbsy,amssymb,graphicx}
\usepackage{times}

\let\mathbf=\boldsymbol
\begin{document}

\title{Edge-Corner Correspondence: \\
Boundary-Obstructed Topological Phases with Chiral Symmetry}
\author{Motohiko Ezawa}
\affiliation{Department of Applied Physics, University of Tokyo, Hongo 7-3-1, 113-8656,
Japan}

\begin{abstract}
The bulk-edge correspondence characterizes topological insulators and
superconductors. We generalize this concept to the bulk-corner
correspondence and the edge-corner correspondence in two dimensions. In the 
\textit{bulk-corner} (\textit{edge-corner}) correspondence, the topological
number is defined for the \textit{bulk} (\textit{edge}), while the
topological phase is evidenced by the emergence of zero-energy \textit{corner%
} states. It is shown that the boundary-obstructed topological phases
recently proposed are the edge-corner-correspondence type, while the
higher-order topological phases are classified into the
bulk-corner-correspondence type and the edge-corner-correspondence type. We
construct a simple model exhibiting the edge-corner correspondence based on
two Chern insulators having the $s$-wave, $d$-wave and $s_{\pm }$-wave
pairings. It is possible to define topological numbers for the edge
Hamiltonians, and we have zero-energy corner states in the topological
phase. The emergence of zero-energy corner states is observable by measuring
the impedance resonance in topological electric circuits.
\end{abstract}

\maketitle

Topological insulators and superconductors are one of the most studied
fields in this decade\cite{Hasan,Qi,Flen,Beenakker}. A topological number is
defined for the bulk. The bulk gap must close at a topological phase
transition point because the topological number cannot change its quantized
value without gap closing. Consequently, the topological phase is evidenced
by the emergence of gapless edge states although the bulk is gapped. This
phenomenon is well known as the bulk-edge correspondence.

We generalize this concept to the bulk-corner correspondence and the
edge-corner correspondence. For definiteness we consider two-dimensional
systems, although generalization to higher dimensions is straightforward\cite%
{SM1}. In the \textit{bulk-corner} (\textit{edge-corner}) correspondence,
the topological number is defined for the \textit{bulk} (\textit{edge}),
while the topological phase is evidenced by the emergence of zero-energy 
\textit{corner} states. There is a case where the topological number for the
bulk is expressed in terms of two topological numbers for the two edges. It
is proper to regard such a system to belong to the
edge-corner-correspondence type. A typical example is given by systems
having square lattice structure\cite{Science}.

Here we summarize the properties whether the gap is open or closed in the
trivial and topological phases in the system subject to the bulk-edge,
bulk-corner and edge-corner correspondences as follows: 
\begin{align}
& \text{bulk-edge}\text{:} & & 
\begin{array}{lccc}
& \text{trivial} & \text{PTP} & \text{topological} \\ 
\text{bulk} & \text{o} & \times & \text{o} \\ 
\text{edge} & \text{o} & \times & \times%
\end{array}%
, \\[10pt]
& \text{bulk-corner}\text{:} & & 
\begin{array}{lccc}
& \text{trivial} & \text{PTP} & \text{topological} \\ 
\text{bulk} & \text{o} & \times & \text{o} \\ 
\text{edge} & \text{o} & \times & \text{o} \\ 
\text{corner} & \text{o} & \times & \times%
\end{array}%
, \\[10pt]
& \text{edge-corner}\text{:} & & 
\begin{array}{lccc}
& \text{trivial} & \text{PTP} & \text{topological} \\ 
\text{bulk} & \text{o} & \text{o} & \text{o} \\ 
\text{edge} & \text{o} & \times & \text{o} \\ 
\text{corner} & \text{o} & \times & \times%
\end{array}%
,
\end{align}%
where o and $\times $ indicate that the gap is open and closed; PTP stands
for the phase-transition point.

It is important to reexamine higher-order topological insulators\cite%
{Fan,Science,APS,Peng,Lang,Song,Bena,Schin,FuRot,EzawaKagome,EzawaPhos,Gei,MagHOTI,Kha}
and superconductors\cite{APS,YanSongWang,WangLiu,WangLin} according to this
classification. Clearly they are classified into these two types: A typical
example of the edge-corner-correspondence type is given by the quadrupole
insulator\cite{Science,Bena}, while a typical example of the
bulk-corner-correspondence type is given by the Kagome lattice\cite%
{EzawaKagome}.

The notion of boundary-obstructed topological phases was recently introduced%
\cite{Khalaf}. Several works on them have succeedingly been reported\cite%
{Asaga,Claes,Tiwari}. It has been proposed that they are realized in
ion-based topological superconductors\cite{Iron}. These phases are
characterized by the property that the bulk-band gap does not close but the
edge-gap closes at the topological phase transition point. Referring to the
properties listed in Eqs.(2) and (3), they must belong to the
edge-corner-correspondence type.

In this paper, we construct a simple model realizing the edge-corner
correspondence and the boundary-obstructed topological phase in order to
make a clear understanding of these phenomena. The model consists of two
Chern insulators with the opposite Chern numbers. We introduce pairing
interactions including the $s$-wave, $d$-wave and $s_{\pm }$-wave pairing.
The system has chiral symmetry. The topological class is BDI, which allows a
topological phase in one dimension but none in two dimensions\cite{Schnyder}%
, Accordingly, a topological phase transition may occur, where the edge gap
is closed while the band gap is open, and zero-energy corner states emerge
in the topological phase. It is possible to define the topological numbers $%
\Gamma _{x}$ and $\Gamma _{y}$ for the edge Hamiltonians along the $x$ and $%
y $ axes, and the topological number $\Gamma _{x}\Gamma _{y}$ for the bulk.
We point out that the edge-corner correspondence is observable by the
impedance measurement in electric circuits.

\textbf{Model Hamiltonian}: We start with the Chern insulator on square
lattice, whose Hamiltonian is given by

\begin{equation}
H_{\text{Chern}}=M\sigma _{z}+\lambda _{x}\sigma _{x}\sin k_{x}+\lambda
_{y}\sigma _{y}\sin k_{y},
\end{equation}%
with Pauli matrices $\sigma _{i}$, spin-orbit interactions $\lambda _{x}$
and $\lambda _{y}$, and%
\begin{equation}
M=t_{x}\cos k_{x}+t_{y}\cos k_{y}-\mu ,
\end{equation}%
where $t_{x}$ and $t_{y}$ are hopping parameters, and $\mu $ is the chemical
potential.

We propose to construct a chiral symmetric model as%
\begin{equation}
H=\tau _{z}H_{\text{Chern}}+\tau _{x}H_{\Delta },  \label{BOTI}
\end{equation}%
where we assume a pairing Hamiltonian\cite{YanSongWang} in the form of%
\begin{equation}
H_{\Delta }=\Delta _{0}+\Delta _{x}\cos k_{x}+\Delta _{y}\cos k_{y},
\end{equation}%
with%
\begin{equation}
\Delta _{x}=\Delta _{s\pm }+\Delta _{d},\qquad \Delta _{y}=\Delta _{s\pm
}-\Delta _{d}.
\end{equation}%
Here, $\Delta _{0}$, $\Delta _{d}$ and $\Delta _{s\pm }$ are the gap
parameters due to the $s$-wave, $d$-wave and $s_{\pm }$-wave pairings\cite%
{Stewart,Hirsch}, respectively.

\textbf{Symmetries:} The Hamiltonian $H$ has chiral symmetry, $\{H,\tau
_{y}\}=0$. The system has time-reversal symmetry, $TH\left( k\right)
T^{\dagger }=H\left( -k\right) $ with $T=\sigma _{x}\tau _{x}K$, where $K$
takes complex conjugation. The system has particle-hole symmetry, $\Xi
H\left( k\right) \Xi ^{\dagger }=-H\left( -k\right) $ with $\Xi =\sigma
_{x}\tau _{z}K$. Hence, the system belongs to the class BDI, where there is
no topological phase in two dimensions\cite{Schnyder}. In addition, the
system has parity symmetry $PH\left( k\right) P^{\dagger }=H\left( -k\right) 
$ with $P=\sigma _{z}$.

\begin{figure}[t]
\centerline{\includegraphics[width=0.49\textwidth]{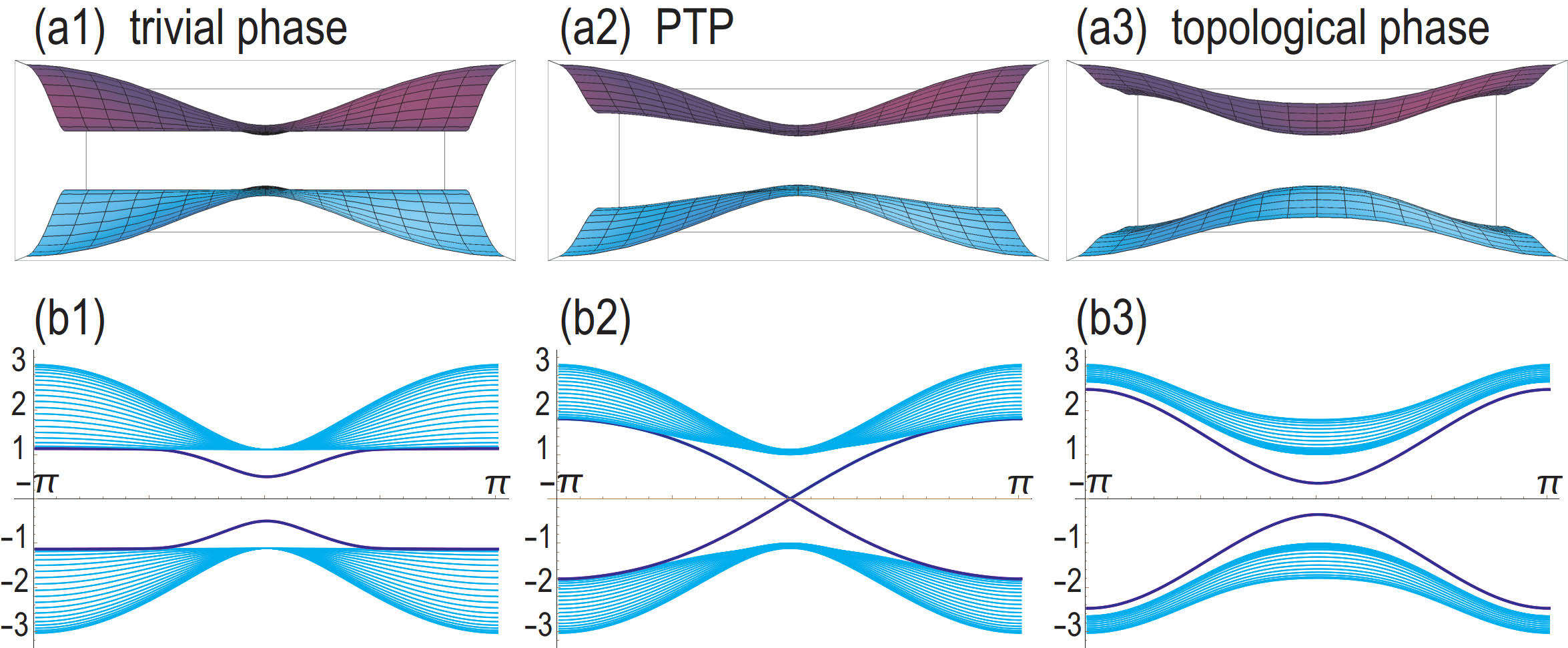}}
\caption{Band structures of (a) the bulk and (b) a nanoribbon. The
nanoribbon bands consist of the bulk bands (in cyan) and the edge bands (in
blue). The edge becomes gapless only at the topological phase transition
point as in (b2), while the bulk gap is always open. We have used parameters 
$\Delta _{d}=0$ for (a1) and (b1), $\Delta _{d}=0.5t$ for (a2) and (b2), and 
$\Delta _{d}=t$ for (a3) and (b3). We have set $t_{x}=t_{y}=\protect\lambda %
_{x}=\protect\lambda _{y}=\protect\mu =t$, $\Delta _{0}=0.5t$ and $\Delta
_{s\pm }=0$. The nanoribbon width is $W=20$.}
\label{FigRibbon}
\end{figure}

\textbf{Bulk:} The energy spectrum of $H$ is given by%
\begin{equation}
E=\pm \sqrt{E_{\text{Chern}}^{2}+E_{\Delta }^{2}},
\end{equation}%
where 
\begin{equation}
E_{\text{Chern}}=\sqrt{M^{2}+\lambda _{x}^{2}\sin ^{2}k_{x}+\lambda
_{y}^{2}\sin ^{2}k_{y}}
\end{equation}%
is the energy of $H_{\text{Chern}}$, and 
\begin{equation}
E_{\Delta }=\Delta _{0}+\Delta _{x}\cos k_{x}+\Delta _{y}\cos k_{y}
\end{equation}%
is the energy of $H_{\Delta }$. The bulk-band gap is 
\begin{equation}
2E\left( 0,0\right) =2\sqrt{\left( t_{x}+t_{y}-\mu \right) ^{2}+\Delta
_{0}^{2}}.  \label{BandGap}
\end{equation}%
The bulk gap $E$ never closes when the $E_{\text{Chern}}$ is gapped.

We show the band structure for various $\Delta _{d}$ in Fig.\ref{FigRibbon}%
(a1)$\sim $(a3) by setting $t_{x}=t_{y}=\lambda _{x}=\lambda _{y}=\mu =t$, $%
\Delta _{0}=0.5t$ and $\Delta _{s\pm }=0$ for simplicity. The bulk gap is
independent of the magnitude of $\Delta _{d}$.

\textbf{Edge theory:} We show the band structure of a nanoribbon for various 
$\Delta _{d}$ in Fig.\ref{FigRibbon}(b1)$\sim $(b3). The edge states become
gapless at a certain phase-transition point as in Fig.\ref{FigRibbon}(b2).
In order to determine this point and to construct the topological phase
diagram, we derive the edge Hamiltonians along the $x$ and $y$ axes.

We first construct the edge Hamiltonian along the $y$ axis. We make a Taylor
expansion at the $\Gamma $ point and decompose the Hamiltonian (\ref{BOTI})
into the unperturbed Hamiltonian $H_{0}$ and the perturbed Hamiltonian $%
H_{1} $, $H=H_{0}+H_{1}$, with%
\begin{align}
H_{0}& =\tau _{z}[\sigma _{z}\left( 2^{-1}t_{x}\partial
_{x}^{2}+t_{x}+t_{y}-\mu \right) -i\lambda _{x}\sigma _{x}\partial _{x}], \\
H_{1}& =\lambda _{y}k_{y}\tau _{z}\sigma _{y}+\left( \Delta _{0}+\Delta
_{x}+\Delta _{y}\right) \tau _{x}+2^{-1}\Delta _{x}\tau _{x}\partial
_{x}^{2},
\end{align}%
where we have omitted insignificant $k_{y}^{2}$ terms\cite{YanSongWang}.
Setting $H_{0}=\tau _{z}H_{0}^{\prime }$, we solve the eigen equation of $%
H_{0}^{\prime }$ as $\psi =\left( \psi _{1},i\eta \psi _{1}\right) $ with $%
\psi _{1}=e^{-\kappa x}e^{ik_{y}y}$, where the penetration depth is 
\begin{equation}
\kappa =\left( \eta \lambda _{x}\pm \sqrt{\lambda _{x}^{2}+2t_{x}m}\right)
/t_{x},
\end{equation}%
with $m=\mu -t_{x}-t_{y}$. By taking an expectation value of $H_{1}$ by $%
\psi $, we obtain the edge Hamiltonian along the $y$ axis as%
\begin{equation}
H_{y}=\eta _{y}\lambda _{y}k_{y}\tau _{z}+\bar{\Delta}_{y}\tau _{x},
\label{EdgeHy}
\end{equation}%
with%
\begin{equation}
\bar{\Delta}_{\alpha }=\Delta _{0}+\Delta _{\alpha }m/t_{\alpha },
\label{EdgeGap}
\end{equation}%
where $\alpha =x,y$; $\eta _{x}=1$ for the lower and $\eta _{x}=-1$ for
upper edges. The edge gap closes at $\bar{\Delta}_{x}=0$.

Similarly we obtain the edge Hamiltonian for the $x$ axis as%
\begin{equation}
H_{x}=\eta _{x}\lambda _{x}k_{x}\tau _{z}+\bar{\Delta}_{x}\tau _{x},
\label{EdgeHx}
\end{equation}%
where $\eta _{y}=1$ for the right and $\eta _{y}=-1$ for left edges. The
edge gap closes at $\bar{\Delta}_{y}=0$.

\begin{figure}[t]
\centerline{\includegraphics[width=0.48\textwidth]{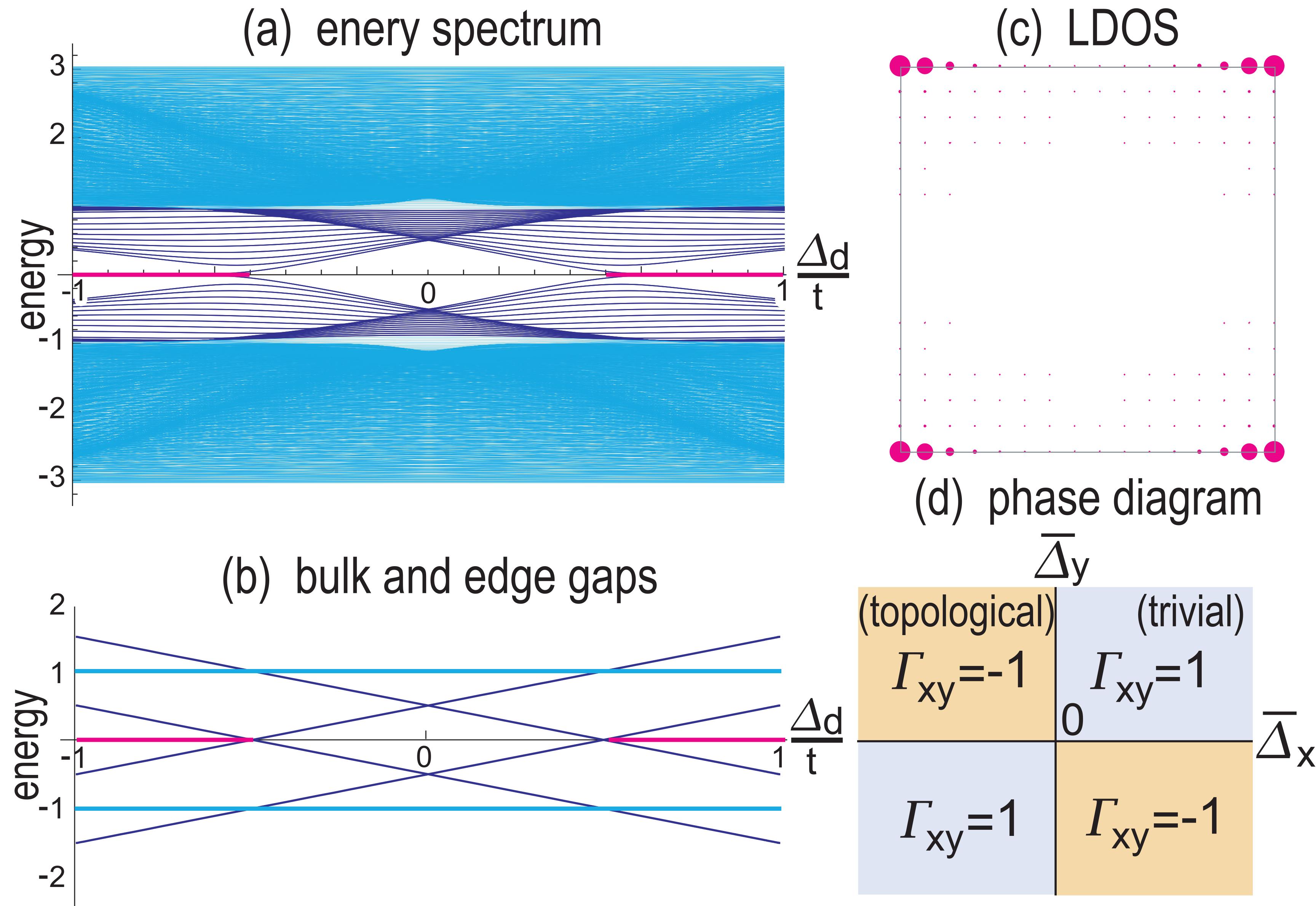}}
\caption{(a) Energy spectrum in square geometry. The bulk and edge bands are
shown in cyan and blue, respectively. The bulk gap is open at the
topological phase transition point. The horizontal axis is $\Delta _{d}/t$,
while the vertical axis is the energy in unit of $t$. We have used a square
with size $N=32$. (b) Bulk gap (in cyan) and edge gap (in blue) as a
function of $\Delta _{d}/t$, which are obtained analytically. The horizontal
lines (in magenta) represent zero-energy corner states. (c) LDOS for
zero-energy corner states. All other parameters are the same as in Fig.1.
(d) Topological phase diagram is determined by the sign of $\bar{\Delta}_{x}%
\bar{\Delta}_{y}$. We can assign the topological number $\Gamma _{xy}$ to
this phase diagram, where $\Gamma _{xy}=1$ ($\Gamma _{xy}=-1$) stands for
the trivial (topological) phase. }
\label{FigCorner}
\end{figure}

\textbf{Edge symmetries: }The edge Hamiltonian $H_{\alpha }$ has chiral
symmetry, $\{H_{\alpha },\tau _{y}\}=0$. The system has time-reversal
symmetry, $TH\left( k\right) T^{\dagger }=H\left( -k\right) $ with $T=\tau
_{x}K$, where $K$ takes complex conjugation. The system has particle-hole
symmetry, $\Xi H\left( k\right) \Xi ^{\dagger }=-H\left( -k\right) $ with $%
\Xi =\tau _{z}K$. Hence, the system belongs to the class BDI, where there is
a topological phase in one dimensions\cite{Schnyder}. In addition, the
system has parity symmetry $PH\left( k\right) P^{\dagger }=H\left( -k\right) 
$ with $P=\tau _{x}$.

\textbf{Edge topological numbers:} The edge chiral index is defined by%
\begin{equation}
\Gamma _{\alpha }=\frac{1}{2\pi i}\text{Tr}\int \tau _{y}H_{\alpha
}^{-1}\partial _{\alpha }H_{\alpha }dk_{\alpha }
\end{equation}%
with $\alpha =x,y$, which is a symmetry-protected-topological number
associated with the chiral symmetry. With the aid of the relation 
\begin{equation}
H_{\alpha }^{-1}\partial _{\alpha }H_{\alpha }=\frac{-2i\bar{\Delta}_{\alpha
}}{\lambda _{\alpha }^{2}k_{\alpha }^{2}+\bar{\Delta}_{\alpha }^{2}},
\end{equation}%
it is calculated as%
\begin{equation}
\Gamma _{\alpha }=-\eta _{\alpha }\text{sgn}\left[ \lambda _{\alpha }\bar{%
\Delta}_{\alpha }\right] .  \label{TopoIndex}
\end{equation}%
We can differentiate the trivial and topological phases by the conditions $%
\bar{\Delta}_{\alpha }>0$ and $\bar{\Delta}_{\alpha }<0$, respectively.

\textbf{Corner theory}: We calculate numerically the energy spectrum and
show it as a function of $\Delta _{d}$ for square geometry in Fig.\ref%
{FigCorner}(a). It consists of the bulk and edge states shown in blue and
cyan, respectively. On the other hand, we can derive analytically the
bulk-band gap as in Eq.(\ref{BandGap}) and the edge-band gap as in Eq.(\ref%
{EdgeGap}), which we show in Fig.\ref{FigCorner}(b). The numerical and
analytical results agree very well. Additionally, we have shown zero-energy
corner states in magenta in Figs.\ref{FigCorner}(a) and (b), which we derive
analytically in what follows. It is prominent that the edge gap closes only
at the phase transition point ($\bar{\Delta}_{\alpha }=0$). We show the
local density of states (LDOS) for the zero-energy corner states in Fig.\ref%
{FigCorner}(c).

We study the corner states analytically. They are described by the
Jackiw-Rebbi solutions\cite{JR} of the edge Hamiltonians. The zero-energy
solution along the $x$-axis is derived by solving (\ref{EdgeHx}), or%
\begin{equation}
H_{x}=-i\eta _{x}\lambda _{x}\tau _{z}\partial _{x}+\bar{\Delta}_{x}\tau
_{x}.
\end{equation}%
The solution is obtained as $\psi _{x}=\left( \psi _{x1},i\eta _{x}\psi
_{x1}\right) $ with%
\begin{equation}
\psi _{x}\left( x\right) =\exp \left[ \pm (\eta _{x}/\lambda _{x})\bar{\Delta%
}_{x}x\right] .
\end{equation}%
Similarly, the zero-energy solution along the $y$-axis reads%
\begin{equation}
\psi _{y}\left( y\right) =\exp \left[ \mp (\eta _{y}/\lambda _{y})\bar{\Delta%
}_{y}y\right] .
\end{equation}%
The wave functions are well defined when they converge, leading to the
condition,%
\begin{equation}
\bar{\Delta}_{x}\bar{\Delta}_{y}<0.  \label{CondiJR}
\end{equation}%
The zero-energy corner states emerge only in this case.

\textbf{Topological phase diagram}: We construct the topological phase
diagram in the ($\bar{\Delta}_{x}$, $\bar{\Delta}_{y}$) plane. As we have
just derived, the condition for the emergence of the zero-energy corner
states is given by (\ref{CondiJR}), or $\bar{\Delta}_{x}\bar{\Delta}_{y}<0$,
where the system is topological. On the other hand, the system is trivial
for $\bar{\Delta}_{x}\bar{\Delta}_{y}>0$, because there are no zero-energy
states. The topological phase diagram is determined by these conditions as
in Fig.\ref{FigCorner}(d), with the phase boundary being given by $\bar{%
\Delta}_{x}\bar{\Delta}_{y}=0$.

We may define the topological number for the bulk, which reproduces the
phase diagram determined in terms of $\bar{\Delta}_{x}\bar{\Delta}_{y}$.
With the use of $\Gamma _{\alpha }$ defined by (\ref{TopoIndex}), it is
given by%
\begin{equation}
\Gamma _{xy}=\Gamma _{x}\Gamma _{y},
\end{equation}%
where $\Gamma _{xy}=1$ for the trivial phase and $\Gamma _{xy}=-1$ for the
topological phase as in Fig.\ref{FigCorner}(d).

\textbf{Electric-circuit realization:} Electric circuits provide us with an
ideal playground to realize various topological phases\cite%
{TECNature,ComPhys,Hel,Lu,Research,Zhao,EzawaTEC,Garcia,Hofmann,EzawaMajo,Tjunc}%
. Especially, higher-order topological phases\cite%
{TECNature,EzawaTEC,EzawaLCR,EzawaSkin} are simulated based on electric
circuits. Topological corner states are observed by impedance resonance\cite%
{TECNature}.

Electric-circuit realization is based on the correspondence\cite%
{TECNature,ComPhys} between the Hamiltonian $H$ and the circuit Laplacian $J$
such that $J=i\omega H$. The Hamiltonian $H_{\text{Chern}}$ is already
discussed in electric circuits\cite{EzawaMajo}. The circuit corresponding to
Hamiltonian $H$ can be constructed in a similar manner. The capacitance
contribute to $i\omega C$, while the inductance contributes to $1/i\omega L$
in the circuit Laplacian. It corresponds to the positive and negative
hoppings in the Hamiltonian. The imaginary hoppings are represented by the
operational amplifiers\cite{Hofmann}. We tune the frequency $\omega $ to be
the critical frequency $\omega _{0}=1/\sqrt{LC}$ so that the circuit
Laplacian is identical to the Hamiltonian.

\begin{figure}[t]
\centerline{\includegraphics[width=0.49\textwidth]{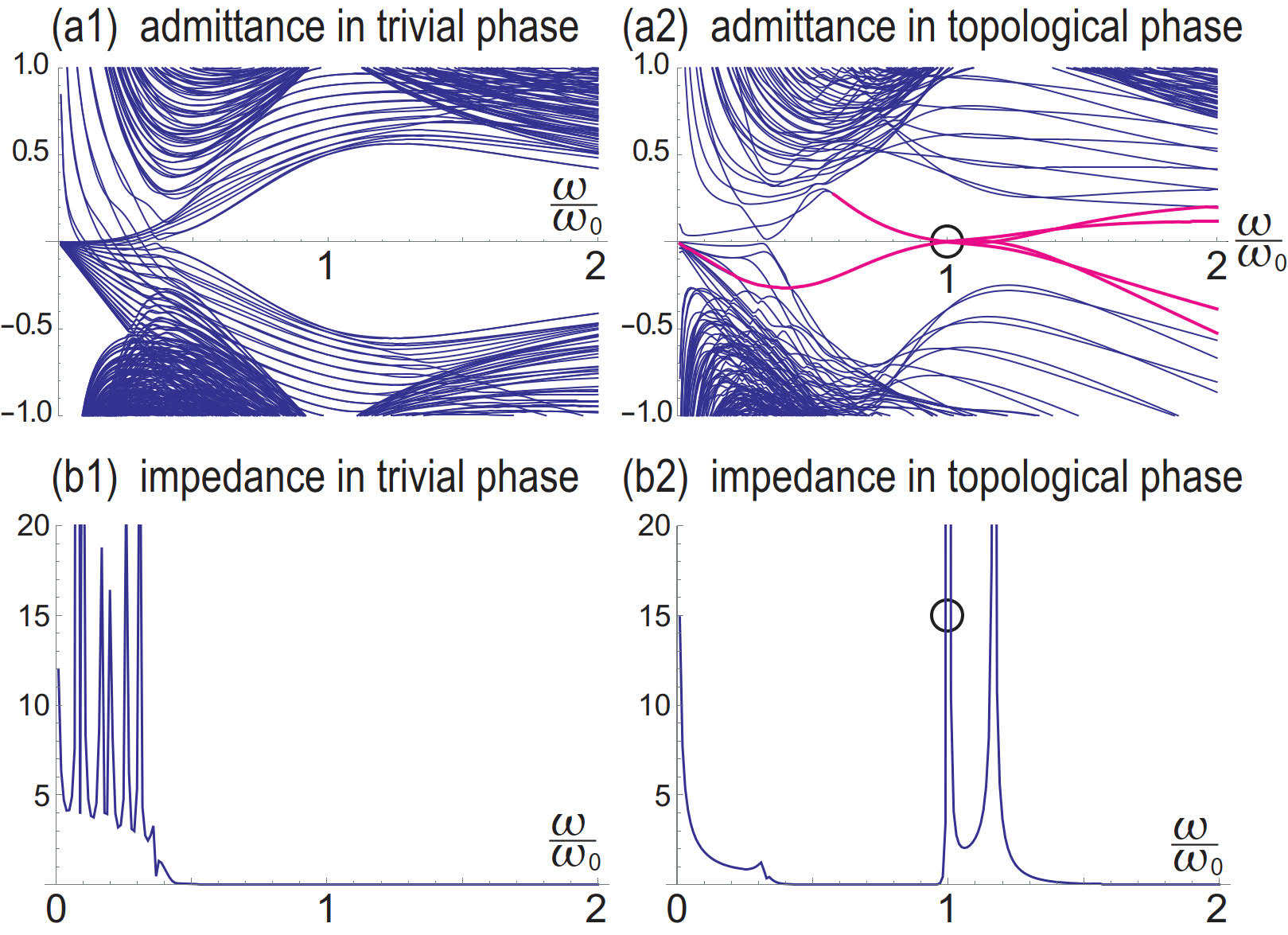}}
\caption{(a) Admittance spectrum in square geometry (a1) for the trivial
phase and (a2) for the topological phase. The horizontal axis is $\protect%
\omega /\protect\omega _{0}$. There are four-fold degenerate zero-admittance
states at $\protect\omega _{0}=\protect\omega $ in the topological phase.
They are marked by a black circle in (a2). (b) Impedance as a function of $%
\protect\omega $ (b1) in the trivial phase and (b2) in the topological
phase. There is a strong peak at $\protect\omega =\protect\omega _{0}$ in
the topological phase, as marked by a black circle in (a2). The horizontal
axis is $\protect\omega /\protect\omega _{0}$.We have used a square with
size $N=16$.}
\label{FigAdmi}
\end{figure}

Admittance is obtained by the eigenvalue of the circuit Laplacian\cite{Hel},
which corresponds to the eigen energy of the Hamiltonian. We show the
admittance spectrum as a function of $\omega $ in Fig.\ref{FigAdmi}(a). The
zero-admittance state appears in the topological phase, while it is absent
in the trivial phase.

The circuit Laplacian is explicitly given by

\begin{equation}
J=\left( 
\begin{array}{cccc}
f_{1} & g_{1} & h & 0 \\ 
g_{2} & f_{2} & 0 & h \\ 
h & 0 & f_{2} & g_{2} \\ 
0 & h & g_{1} & f_{1}%
\end{array}%
\right)
\end{equation}%
with%
\begin{align}
f_{1}& =i\omega \left( C_{x}\cos k_{x}+C_{y}\cos k_{y}\right) -i\omega
C_{x}-i\omega C_{y}+i\omega C_{m},  \notag \\
f_{2}& =(i\omega L_{x})^{-1}\cos k_{x}+(i\omega L_{y})^{-1}\cos k_{y}  \notag
\\
& -(i\omega L_{x})^{-1}-(i\omega L_{y})^{-1}+(i\omega L_{m})^{-1},  \notag \\
g_{1}& =i\omega C_{X}e^{ik_{y}}+(i\omega L_{X})^{-1}e^{-ik_{y}}+R^{-1}\left(
e^{ik_{x}}-e^{-ik_{x}}\right) ,  \notag \\
g_{2}& =(i\omega L_{X})^{-1}e^{ik_{y}}+i\omega C_{X}e^{-ik_{y}}+R^{-1}\left(
e^{ik_{x}}-e^{-ik_{x}}\right) ,  \notag \\
h& =i\omega C_{\Delta 0}+i\omega C_{\Delta x}\cos k_{x}+i\omega C_{\Delta
y}\cos k_{y}.
\end{align}%
It is straightforward to design electric circuits corresponding to this
circuit Laplacian.

We analyze the impedance\cite{Hel}, which is defined by $%
Z_{ab}=V_{a}/I_{b}=G_{ab}$ where $G=J^{-1}$ is the Green function. It
diverges at the frequency where the admittance is zero ($J=0$). Note that
the impedance and the admittance have inverse relation. Taking the nodes $a$
and $b$ at two corners, we show the impedance in the trival and topological
phases in Figs.\ref{FigAdmi}(b1) and (b2), respectively. A strong impedance
peak is observed at the critical frequency $\omega _{0}$ in the topological
phase, while it is absent in the trivial phase. It signals the emergence of
zero-energy corner states. Hence, the edge-corner correspondence is
observable in electric circuits.

\textbf{Discussions:} We have presented a classification of various
higher-order topological phases based on the bulk-corner and edge-corner
correspondences. We have constructed a simple model exhibiting the
edge-corner correspondence, where topological numbers are defined for edges.
It provides us with a clear picture of the boundary-obstructed topological
phase transition. Furthermore, we have argued that the emergence of
topological corner states is detectable by observing impedance peaks in
topological electric circuit. The merit of electric circuits is that we can
tune the values of elements relatively easily, which enables to realize a
topological phase transition.

The generalization to three dimensions is straight forward. We may consider
bulk-hinge, surface-hinge correspondences for the second-order topological
phases\cite{SM1}. We may also consider bulk-corner, surface-corner and
hinge-corner correspondences for the third-order topological phases\cite{SM1}%
.

The author is very much grateful to N. Nagaosa for helpful discussions on
the subject. This work is supported by the Grants-in-Aid for Scientific
Research from MEXT KAKENHI (Grants No. JP17K05490 and No. JP18H03676). This
work is also supported by CREST, JST (JPMJCR16F1).

\clearpage\newpage
\onecolumngrid
\def\theequation{S\arabic{equation}}
\def\thefigure{S\arabic{figure}}
\def\thesubsection{S\arabic{subsection}}
\setcounter{figure}{0}
\setcounter{equation}{0}

\centerline{\textbf{\Large Supplemental Material}}\bigskip 

\centerline{\large\textbf{Edge-Corner Correspondence: Boundary-Obstructed Topological Phases with Chiral
Symmetry}} \medskip \centerline{Motohiko Ezawa} 
\centerline{Department of
Applied Physics, University of Tokyo, Hongo 7-3-1, 113-8656, Japan}

\subsection{Second-order topological phases in three dimensions}

The second-order topological phases in three dimensions are characterized by
the phenomena that gapless hinge states emerge although the bulk and the
surface are gapped. They are classified into the bulk-hinge-correspondence
type and the surface-hinge-correspondence type according to the geometrical
object upon which the topological numbers are defined. Hereafter, let us
abbreviate the A-B-correspondence type as the A-B type. (i) In the
bulk-hinge type, the bulk gap closes at the phase transition point. The
topological number is defined for the bulk. (ii) In the surface-hinge type,
the surface gap closes but the bulk gap remains to open at the phase
transition point. The topological number is defined for the surface.

These phenomena are summarized as

\begin{align}
& \text{(i) bulk-hinge correspondence:} & & 
\begin{array}{lccc}
& \text{trivial} & \text{PTP} & \text{topological} \\ 
\text{bulk} & \text{o} & \times & \text{o} \\ 
\text{surface} & \text{o} & \times & \text{o} \\ 
\text{hinge} & \text{o} & \times & \text{o}%
\end{array}%
, \\[10pt]
& \text{(ii) surface-hinge correspondence:} & & 
\begin{array}{lccc}
& \text{trivial} & \text{PTP} & \text{topological} \\ 
\text{bulk} & \text{o} & \text{o} & \text{o} \\ 
\text{surface} & \text{o} & \times & \text{o} \\ 
\text{hinge} & \text{o} & \times & \text{o}%
\end{array}%
.
\end{align}

\subsection{Third-order topological phases in three dimensions}

The third-order topological phases in three dimensions are characterized by
the phenomena that zero-energy corner states emerge although the bulk, the
surface and the hinge are gapped. They are classified into three types, the
bulk-corner type, surface-corner type and the hinge-corner type, according
to the geometrical object upon which the topological numbers are defined.
(i) In the bulk-corner type, the bulk gap closes at the phase transition
point. The topological number is defined for the bulk. (ii) In the
surface-corner type, the surface gap closes but the bulk gap remains to open
at the phase transition point. The topological number is defined for the
surface. (iii) In the hinge-corner type, the hinge gap closes but the bulk
and surface gaps remain to open at the phase transition point. The
topological number is defined for the hinge.

These phenomena are summarized as 
\begin{align}
& \text{(i) bulk-corner correspondence:} & & 
\begin{array}{lccc}
& \text{trivial} & \text{PTP} & \text{topological} \\ 
\text{bulk} & \text{o} & \times & \text{o} \\ 
\text{surface} & \text{o} & \times & \text{o} \\ 
\text{hinge} & \text{o} & \times & \text{o} \\ 
\text{corner} & \text{o} & \times & \times%
\end{array}%
, \\[10pt]
& \text{(ii) surface-corner correspondence:} & & 
\begin{array}{lccc}
& \text{trivial} & \text{PTP} & \text{topological} \\ 
\text{bulk} & \text{o} & \text{o} & \text{o} \\ 
\text{surface} & \text{o} & \times & \text{o} \\ 
\text{hinge} & \text{o} & \times & \text{o} \\ 
\text{corner} & \text{o} & \times & \times%
\end{array}%
, \\[10pt]
& \text{(iii) hinge-corner correspondence:} & & 
\begin{array}{lccc}
& \text{trivial} & \text{PTP} & \text{topological} \\ 
\text{bulk} & \text{o} & \text{o} & \text{o} \\ 
\text{surface} & \text{o} & \text{o} & \text{o} \\ 
\text{hinge} & \text{o} & \times & \text{o} \\ 
\text{corner} & \text{o} & \times & \times%
\end{array}%
.
\end{align}

\end{document}